\documentclass[12pt]{article}
\usepackage{graphicx}
\usepackage{epstopdf}  

\usepackage{lineno}

\usepackage{natbib}

\begin{document}

\begin{center}
{\large Clustering of variables for enhanced interpretability of predictive models.}

\vskip 3mm

Evelyne Vigneau
\end{center}

\noindent StatSC, Oniris, INRA,

\noindent 44 322 NANTES Cedex 03, FRANCE

\noindent evelyne.vigneau@oniris-nantes.fr

\vskip 3mm

\begin{abstract}
	A new strategy is proposed for building easy to interpret predictive models in the context of a high-dimensional dataset, with a large number of highly correlated explanatory variables. The strategy is based on a first step of variables clustering using the CLustering of Variables around Latent Variables (CLV) method. The exploration of the hierarchical clustering dendrogram is undertaken in order to sequentially select the explanatory variables in a group-wise fashion. For model fitting implementation, the dendrogram is used as the base-learner in an L2-boosting procedure. The proposed approach, named \textit{lmCLV}, is illustrated on the basis of a toy-simulated example when the clusters and predictive equation are already known, and on a real case study dealing with the authentication of orange juices based on $^1$H-NMR spectroscopic analysis. In both illustrative examples, this procedure was shown to have similar predictive efficiency to other methods, with additional interpretability capacity. It is available in the R package \texttt{ClustVarLV}.
\end{abstract}

\section{Introduction}
\label{intro}

In the context of high-dimensional data with a large number of variables, \emph{p}, and small number of observations, \emph{n}, such as microarray data, metabolomic and volatolomic data  (among a large variety of -omic data collected using high-throughput fingerprinting technologies), most recent statistical modelling strategies aim to achieve both efficient prediction and enhanced interpretability outcomes. For 20 years, the question of variable selection has driven a great deal of work, from discrete processes in which variables are either retained or removed, to lasso regularization processes in which several model coefficients may be shrunken towards zero \citep{Hastie09, Tib96}. Regularization strategies discussions are still an active topic due to the high-dimensional of current data, either with model-based approach as in \citet{Celeux19} or in combination with filtering step as in \citet{Algamal19}.

However, as Lasso tends to arbitrarily select one predictive variable among a group of highly correlated relevant variables, the sparsity principle of Lasso have been embedded into strategies for the selection of grouped predictors. Fused Lasso \citep{Tib05} and Group Lasso \citep{Yuan06} are designed for explanatory variables naturally ordered (as in vibrational spectroscopy) or arranged into groups (as for design factors). Sparse Group Lasso \citep{Fried10, Zeng17} yields sparsity at both the group and individual feature levels, in order to select groups and predictors within a group. The Octagonal Shrinkage and Clustering Algorithm for Regression (OSCAR) \citep{Bond08} is another approach which combines an $L_1$-type penalty and a pairwise $L_\infty$-type penalty on the model's coefficients. Thus OSCAR allows to simultaneously select variables and perform supervised clustering in the context of linear regression. The CLERE methodology \citep{Yengo16} is also based on the clustering of the regression coefficients using a Gaussian latent mixture model.  Several of the penalized least squares approaches have Bayesian analogues also developed for variable selection and grouping \citep{Curtis11, Chak19}. 

The model we propose here stems from another family of approaches. Emphasis is put on reducing the dimensionality of the data when a large number of highly correlated explanatory variables ($p \gg n$) has been collected. We consider reduction dimension approaches that define latent components as being linear combinations of the explanatory variables and taking their correlation-structure into account. It is also desirable that these latent components are easy to interpret by making several component loadings to be exactly zeros for instance. This goal has also been addressed with approaches designed to constrain Principal Component loadings to be shrunk to zeros \citep{Jolli03, Chip05, Cox18}, or by first performing a cluster analysis in order to construct interpretable components \citep{Enki13, Buhl13}.

Our proposal is also to first perform a clustering of the explanatory variables and then to fit a linear model between a response variable $Y$ and $p$ explanatory variables $X_j$ ($j=1,\ldots,p$) in a groupwise fashion. The algorithm adopted herein is easy to implement and is oriented towards the interpretability of the latent component introduced sequentially into the model. In the case study considered herein, each selected latent component can be assumed to be associated with a compound from its chemical spectrum in NMR (Nuclear Magnetic Resonance). With metabolomic data, it is possible to imagine that the latent components associated to subsets of metabolites may be related to specific biological pathways. Within the context of DNA microarray data, similar studies have been undertaken \citep{Hastie01, Park07} in which hierarchical cluster analysis allows to identify supergenes, obtained by averaging genes within the clusters. These supergenes are thereafter used to fit regression models, thereby attaining concise interpretation and accuracy. One of the main differences in this study, compared with previous research works \citep{Hastie01, Park07}, is that representatives of the clusters of variables, or latent variables, are not necessarily an average of the observed variables, but can be components that best reflect the variability within each cluster of variables. The second difference is in the procedure adopted for the progressive construction of the regression linear model. 

The clustering of the explanatory variables is based herein on the CLV (CLustering of Variables around Latent Variables) approach \citep{Vig03, Vig16}, implemented in the \texttt{ClustVarLV} R package \citep{Vig15}. In summary, the CLV approach consists of clustering together highly correlated variables into clusters (or groups of variables), while exhibiting within each cluster a latent variable (or latent component) representative to this cluster. It turns out that each latent component is defined as a linear combination of only the variables belonging to the corresponding cluster. From this point of view, CLV components are sparse components in the space of observations, and aim to best reflect the variance-covariance structure between the explanatory variables. Using the same approach, but defining a slightly different criterion, \citet{Enki13} also proposed a clustering of variables method for producing interpretable principal components. \citet{Figueiredo15} investigated an approach for clustering of variables, based on the identification of a mixture with bipolar Watson components defined on the hypersphere. Herein, the central feature of the clustering procedure is a bottom-up aggregation approach of the explanatory variables involving the CLV criterion.

The dendrogram obtained is then explored in order to identify and select the predictive group's latent components regarding the response variable $Y$. In contrast with a previous study \citep{Chen16}, hierarchical clustering, which is the more time consuming step, is performed only once and the fitting model stage has been modified accordingly. More precisely, an L2-boosting procedure for which the base-learner model is the CLV hierarchical algorithm has been considered. It consists to, iteratively, select a group of explanatory variables. The residuals of the response variable is then regressed on the latent component associated with the selected group and the predicted response is updated with the shrunken version of this local predictor.

The proposed methodology combining variables clustering and iterative linear model fitting, designated as \textit{lmCLV}, is described in Section \ref{Sec:methodo}. Section \ref{Sec:appli} includes a simple simulated dataset in order to illustrate the behavior of the procedure in a known context, as well as one real case study dealing with the authentication of orange juice based on $^1$H-NMR spectroscopy.

\section{Methodology} \label{Sec:methodo}

\subsection{Notation} \label{Sec:nota}

We consider the high-dimensional linear model :
\begin{equation} \label{Eq:linearmodel}
\mathbf{y} = \mathbf{X} \, \mathbf{\beta} + \mathbf{\varepsilon},
\end{equation}
where $\mathbf{y}=[y_{i}]$ is a $n \times 1$-dimensional random response vector;  $\mathbf{X}=[\mathbf{x}_1 \vert \ldots \vert \mathbf{x}_p]=[x_{ij}]$, a $n \times p$-dimensional quantitative explanatory variables matrix (with $i = 1, \ldots, n$ and $j = 1, \ldots, p$); $\mathbf{\beta}$, a $p \times 1$ regression coefficients vector and $\mathbf{\varepsilon}$ the $n \times 1$-dimensional random vector of the residuals, with zero mean and constant variance. 

We consider contexts where $ p > n$ or $p \gg n$ and where the explanatory variables may be arranged into groups of highly correlated variables.

Both the explanatory variables matrix and the response vector are assumed to be column-centered. In addition, the user may choose to standardize, or not, the variables to a unit variance.

\subsection{\textit{lmCLV}'s outlines} \label{Sec:outlines}

\textit{lmCLV} combines two main methods: \begin{enumerate}
	\item The CLV method which is performed first.The similarities between the variables, herein evaluated on the basis of their covariance or correlation coefficients, are depicted by a tree diagram which is used as the learner for the model fitting method. The clustering of variables, using CLV method, is detailed in Sect.\ref{Sec:clust}.
	\item The L2-boosting procedure which provides an efficient iterative model fitting method. The outline of the L2-boosting algorithm is depicted in Sect.\ref{Sec:model}, and the way the CLV base-learner is used in the course of this procedure is explained in Sect.\ref{Sec:baselearner}.
\end{enumerate}

\subsection{Clustering of the variables} \label{Sec:clust}

The clustering of the explanatory variables using the CLV method for directional groups \citep{Vig03, Vig15} aims to both identify clusters of variables and define a latent component associated with each cluster.

For a given number of clusters, $K$, the aim of the CLV method is to seek a partition $\mathcal{G_K}=\{G_1, \ldots G_K\}$ of the variables into $K$ disjoint clusters and a matrix $\mathbf{C}_K=[\mathbf{c}_1 \vert \ldots \vert \mathbf{c}_K]$ of $K$ latent variables, each being associated with one cluster, so as to maximize the internal coherence within the clusters. When the agreement between the variables is assessed using their covariance or correlation coefficients, regardless of the sign of these coefficients, the clusters we are looking for are named directional groups. In this case, the aim is to define the partition, $\mathcal{G_K}$, and group's latent variables matrix, $\mathbf{C}_K$, so that to get the optimal value $T^*$ of the CLV criterion, so that: 
\begin{equation} \label{Eq:crit}
T^* := \max_{(\mathcal{G_K},\mathbf{C_K})} \sum_{k=1}^K\sum_{j=1}^p \delta_{kj} \; cov(\mathbf{x}_j, \mathbf{c}_k)^2,
\end{equation}
under the constraints $||\mathbf{c}_k||=1$.

In eq.(\ref{Eq:crit}), $\delta_{kj} = 1$ if the $j^{th}$ variable belongs to the group $G_k$, and $\delta_{kj} = 0$ otherwise. In other terms, $(\delta_{kj})$ is the generic term of a binary matrix $\mathbf{\Delta}$ of the group's membership of the $p$ variables. $\mathbf{\Delta}$ has only one nonzero element per row.

It is easy to show \citep{Vig03} that when the partition $\mathcal{G_K}$ is fixed, the latent variable in a cluster $G_k \, (k=1,...,K)$ is defined as the first standardized principal component of a data matrix formed by the variables belonging to this cluster. The latent variable $\mathbf{c}_k$ associated with the cluster $G_k \,(k=1,...,K)$ can therefore be expressed as a linear combination of the variables of this group:

\begin{equation} \label{Eq:ck}
\mathbf{c}_k= \sum_{j/ \delta_{kj}=1} v_j \mathbf{x}_j.
\end{equation}

\medskip

For various numbers of clusters, from $K=p$ clusters, in which each variable is considered to form a cluster by itself, to $K=1$, where there is a single cluster containing all the variables, an ascendant hierarchical clustering algorithm is also proposed with respect to the CLV criterion and aggregating rules detailed in \citet{Vig03}. The strategy proposed herein consists of firstly constructing the whole hierarchy of the $p$ explanatory variables and to explore repeatedly the dendrogram obtained. 

\subsection{L2-boosting procedure for model fitting}\label{Sec:model}

The second feature of the strategy is a boosting procedure, which can also be represented as a functional gradient descent (FGD) algorithm \citep{Efron16,Buhl07}, performed on the residuals of an iteratively updated shrunken linear model for the prediction of the response variable $\mathbf{y}$. The outline of the algorithm can be depicted as follows: 

\begin{description}
	\item[1.] Set $m=0$, and initialize $\mathbf{\hat{y}}^{(0)}$, for instance by choosing $\mathbf{\hat{y}}^{(0)} = \mathbf{1_n} \bar{y}$,  with $\mathbf{1_n}$ a vector of ones of length $n$, and $\bar{y}= 1/n \sum_{i=1}^{n} y_i$,	\vspace{0.1cm}
	\item[2.] Increase $m$ by 1, and compute the residuals $e_i^{(m)}=y_i - \hat{y_i}^{(m-1)}$ for $i=1, \ldots, n$;
	\vspace{0.1cm}
	\item[3.] Apply the base-learner procedure to the actual residuals. The aim at this step is to identify the "best" CLV latent component denoted $\mathbf{c}^{*(m)}$. This base-learner procedure will be described in Sect.\ref{Sec:baselearner};
	\vspace{0.1cm}
	\item[4.] Update the predictive function, i.e. $\mathbf{\hat{y}}^{(m)} = \mathbf{\hat{y}}^{(m-1)} + \nu \, \alpha^{(m)} \, \mathbf{c}^{*(m)}$ where $0 < \nu \leq 1$ is a shrinkage parameter and $\alpha^{(m)}$ the Ordinary Least Squares (OLS) coefficient estimate of the linear regression of $\mathbf{c}^{*(m)}$ on $\mathbf{e}^{(m)}=[e_1^{(m)}, \ldots, e_n^{(m)}]$.
	\vspace{0.1cm}
	\item[5.] Return to step 2., until $m=M$ ($M$ being a large predetermined integer).
\end{description}

This procedure depends on two parameters: the stopping iteration parameter, $M$, and the shrinkage parameter, $\nu$, which can be determined via cross-validation or other information criterion as previously suggested in \citet{Buhl07}.

In practice, because our base-learner procedure returns one-dimensional components associated with sequential group-wise selections of the explanatory variables, we have often observed (as it will be shown in Sect.\ref{Sec:appli}) that a relatively high value of $\nu$ (greater than 0.5) generally performs better. Moreover, as the predictive ability of the model appeared to be relatively stable, large values of $M$ (say 50 or more) are not necessarily useful.

\subsection{Base-learner procedure} \label{Sec:baselearner}

The third step of the algorithm presented in Sect.\ref{Sec:model} is the core of the proposed strategy. At each iteration, $m$, of the algorithm, the aim is to select a cluster of explanatory variables and their associated representative, i.e. the group's latent component $\mathbf{c}^{*(m)}$. This choice is guided by CLV hierarchical clustering of the $p$ explanatory predictors of  $\mathbf{X}=[\mathbf{x}_1 \vert \ldots \vert \mathbf{x}_p]$ (Sect.\ref{Sec:clust}).

\begin{description}
	\item[$\bullet$] For each size of partition (from $p$ clusters to one cluster), we first aim to identify the CLV latent component which has the largest correlation (in absolute value) with the actual residuals $\mathbf{e}^{(m)}$. 
	For $q=1, \ldots p$, that is for each partition $\mathcal{G}_{p-q+1}$, we define :
	\begin{equation}
	\mathbf{c}^*_q:=\max_{k\in\{1,\ldots, (p-q+1)\}} |cor(\mathbf{c}_k,\mathbf{e}^{(m)})|.
	\end{equation} 
	
	\item[$\bullet$] The next step consists of choosing a specific level $q$ between 1 and $p$, that is a latent component $	\mathbf{c}^*_q$ and its associated group of predictors $G_q^*$, in such a way that $G_q^*$ is as large as possible while accommodating an undimentionality criterion. The unidimensionality condition is assessed herein using the modified Kaiser-Guttman (KG) rule \citep{Karlis03}.
	
	If we denote $m$ this specific level:
	\begin{equation}
	m:=arg \max_{q\in\{1,\ldots, p\}} (|G_q^*| \, / \, \lambda_1> \mathcal{L} \texttt{ and }\lambda_2 \leq \mathcal{L}),
	\end{equation} 
	where $|G_q^*|$ denotes the cardinal of $G_q^*$, $\lambda_1$ and $\lambda_2$ are respectively the first and the second largest eigenvalues of the correlation matrix of the explanatory variables belonging to  $G_q^*$ and the threshold $\mathcal{L}$ is defined according to \citet{Karlis03} by:
	\[
	\mathcal{L}=1 + 2 \sqrt{\frac{|G_q^*|-1}{p-1}}.
	\]
	
\end{description}

The latent component $\mathbf{c}^{*(m)}$ and the groups of explanatory variables in $G_m^*$ are returned to the main algorithm (Sect.\ref{Sec:model}) which continues at  step \textbf{4}.

Finally, it can be noticed that at each step the base learner returns a latent component which is itself a linear combination of a subset of the explanatory variables (eq.\ref{Eq:ck}). It is therefore easy to reformulate the prediction function in terms of a linear combinations of the $p$ predictors, and to obtain an estimate of the coefficients' vector $\mathbf{\beta}$   (eq.\ref{Eq:linearmodel}).

\section{Applications} \label{Sec:appli}

\subsection{A toy-simulated example} \label{Sec.simul}

The \textit{lmCLV} procedure is illustrated in this section using a simple example based on a simulated model as in \citet{Chen16}. We considered herein $p=70$ explanatory variables supposed to be measured on $n=100$ observations. Moreover, these variables were assumed to be structured into five groups ($\mathcal{G}_1$ to $\mathcal{G}_5$) of various sizes: $\mathcal{G}_1$ was the largest group consisting of 35 variables, $\mathcal{G}_2$ was the smallest group with 5 variables, whereas $\mathcal{G}_3, \mathcal{G}_4$ and $\mathcal{G}_5$ were 10 variables each. 

Each group of variables was generated around a prototypical variable. The five prototypes were centered and standardized random variables with a known structure of covariance. In practice, $n$ realizations of a vector $(\mathcal{Z}_{1},\ldots,\mathcal{Z}_{5})^t$ were generated from a centered multivariate normal distribution with a covariance matrix $\mathbf{\Sigma}$:

\begin{eqnarray}
\mathbf{\Sigma} =
\left( \begin{array}{ccccc}
1   & 0.5 & 0.5 & 0.5 & 0.1  \\
0.5 & 1   & 0.5 & 0.1 & 0.1  \\
0.5 & 0.5 & 1   & 0.1 & 0.1 \\
0.5 & 0.1 & 0.1 & 1   & 0.1 \\
0.1 & 0.1 & 0.1 & 0.1 & 1
\end{array} \right).
\end{eqnarray}

Let us denote $\mathbf{Z}$ the $n$ x 5 generated prototypical matrix. Then, the variables of each group were randomly simulated according to a multivariate normal distribution  $\mathcal{N}(\mathbf{0_n},\mathbf{I_n})$, where $\mathbf{0_n}$ represents the $n$-dimensional null vector and $\mathbf{I_n}$ the $n \times n$ identity matrix, as follows:

\begin{equation} \label{Eq:simulxj}
\mathbf{x}_j = \omega_j \,\mathbf{Z} \mathbf{\Lambda}^t +  \mathbf{\varepsilon}_j  \texttt{   for } j \in \{1,\ldots, p\},
\end{equation}
\noindent where $\mathbf{\Lambda}$ is a $p$ x 5 binary matrix defining the allocation of explanatory variables into the 5 groups. The column-wise marginal sum of $\mathbf{\Lambda}$ was $[35,5,10,10,10]$. In eq.(\ref{Eq:simulxj}), $\omega_j \in \lbrace +1, -1\rbrace $ was used to randomly create positive or negative correlations between each simulated variable $\mathbf{x}_j$ in a group  and the associated prototypical variable, in order to generate directional groups of variables.

Finally the response variable $\mathbf{y}$ ($\in  \textbf{R}^n$) was generated with the following model:
\begin{equation} \label{Eq:simuly}
\mathbf{y}  =  \mathbf{Z} \mathbf{b} + \mathbf{\varepsilon} \texttt{     with      } \textbf{b}=[1,5,3,0,0]^t,
\end{equation}
\noindent where $\varepsilon$ resulting from a multivariate normal distribution $\mathcal{N}(\mathbf{0_n},\mathbf{I_n})$.

The response $\bf y$ was supposed to be mainly related to the variables of the smallest group, $\mathcal{G}_2$, moderately to the variables in $\mathcal{G}_3$ and the smallest for the most numerous variables in $\mathcal{G}_1$. According to the parameters of the simulation, the expected correlation coefficients between $\bf y$ and five prototypical variables are 0.69, 0.96, 0.82, 0.18 and 0.12, respectively. This toy-simulated example was designed to represent a simplified, but realistic, case study. 

\begin{figure}[h]
	\centering
	\includegraphics[width=\linewidth]{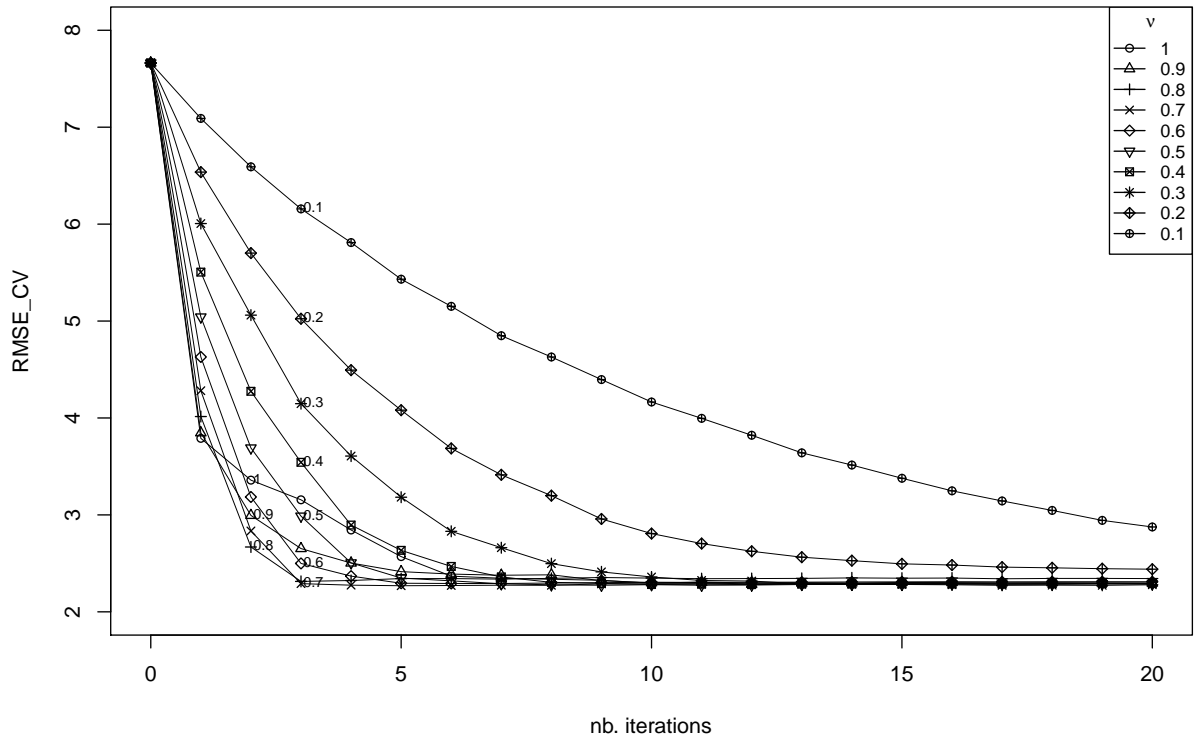}
	\caption{Evolution of the Root Mean Squared Error in Cross-validation with \textit{lmCLV}, according to the number of iterations, and for various values of the shrinkage parameter, $\nu$, (from 1 to 0.1).} 
	\label{Fig:simul_nu}
\end{figure}

The choice of the shrinkage parameter $\nu$ is discussed on the basis of the Root Mean Squared Error criterion evaluated using a five-fold cross-validation procedure, denoted $RMSE_{CV}$. This criterion was assessed at each iteration $m$ from 0 (null model) up to $M=20$. Figure \ref{Fig:simul_nu} shows that as the parameter $\nu$ becomes smaller, the decrease in prediction errors becomes slower. In this example, a value of $\nu=0.7 \texttt{ or } 0.8$ led to a low level of errors ($RMSEP_{CV}=2.29$)  after only three iterations of the algorithm. In the following, $\nu$ was fixed to 0.7.

At the first iteration, and for each fold of the cross-validation procedure, the variable numbers 36 to 40 ($\mathcal{G}_2$) were in the cluster associated with the selected CLV latent component (for one of the five folds the variable 33 was also added, and for another fold the variable 46 was added). At the second iteration, the $\mathcal{G}_3$ variables (41 to 50) were always retained (except for one fold, for which the variable 48 was lacking). Finally, at the third iteration, as expected, all the variables associated with $\mathcal{G}_1$, number 1 to 35, were in the selected cluster of variables.

\begin{figure}[h]
	\centering
	\resizebox*{5cm}{!}{\includegraphics{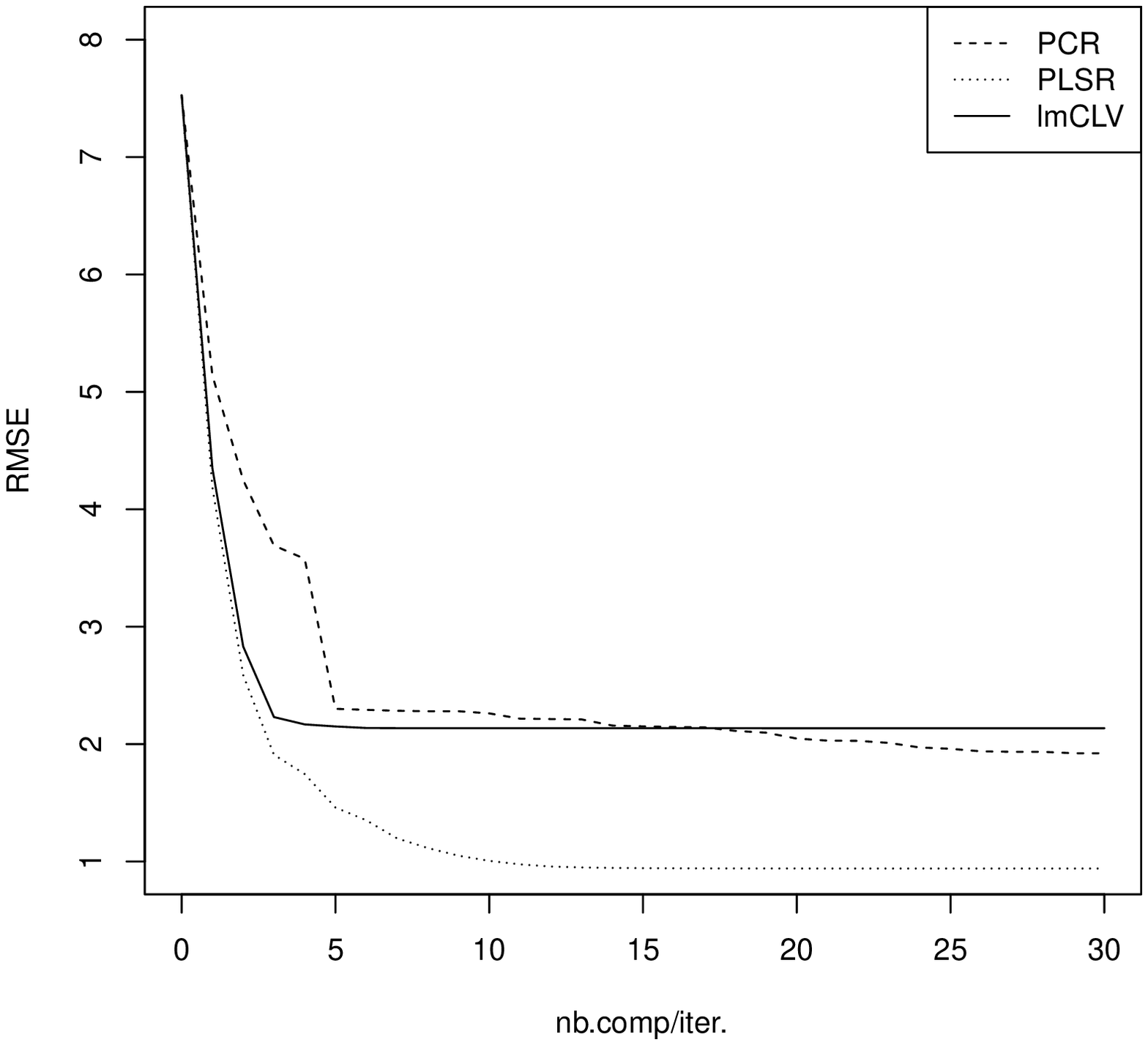}}
	\hspace{5pt}
	\resizebox*{5cm}{!}{\includegraphics{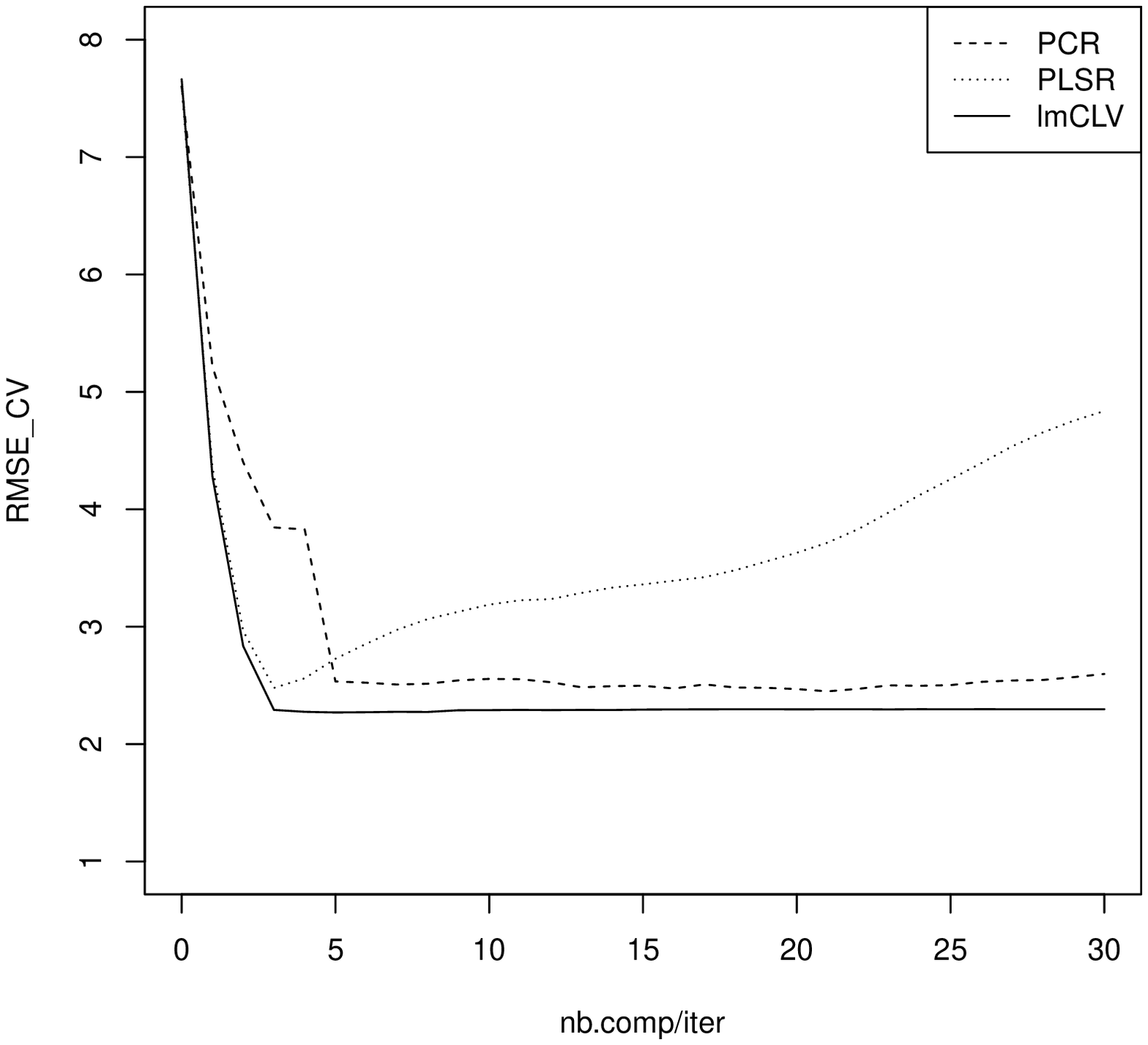}}
	\caption{Evolution of the Root Mean Squared Errors } 
	\label{Fig_simul_compar}
\end{figure}

For comparison of \textit{lmCLV} with two usual regression methods based on latent components, namely Principal Components Regression (PCR) and Partial Least Squares Regression (PLSR), the same dataset was considered, using the same five consecutive segments for Cross-Validation. The evolution of the $RMSE_{CV}$ according to the number of latent components included are shown in Figure \ref{Fig_simul_compar}(b). The left hand side panel (Figure \ref{Fig_simul_compar}(a)) shows the evolution of the errors in prediction, more precisely the $RMSE$, evaluated on the calibration set formed by the whole dataset. As can be observed in Figure \ref{Fig_simul_compar}(a), PLSR and, to a weaker extend, PCR  are prone to over-fitting, with the $RMSE$ value reaching 0.94 when $p=70$ components are included (i.e. the OLS solution). On the contrary the $RMSE$ with \textit{lmCLV} remained relatively stable with increasing numbers of iterations. As a matter of fact, the higher the number of iterations, the more often the same groups of variables, and thus the same latent components, are likely to be included in the model. However, the loadings of these repeated latent components are becoming increasingly smaller, which has no impact on the overall quality of the model. With three latent components, extracted during three iterations of the algorithm, \textit{lmCLV} made it possible to obtain a value of $RMSE=2.23$ and of $RMSE_{CV}=2.29$. Regarding the $RMSE_{CV}$ criterion, in this example, \textit{lmCLV} performed a little better than PLSR and PCR. As with \textit{lmCLV}, the PLSR solution with three components could be retained. However, except for the first PLS component which was quite well correlated ($r=-0.94$) with the third latent component retained with \textit{lmCLV} (also well correlated with the third prototypical variable), the other components were not pairwise well correlated. In fact, the PLS components are two by two uncorrelated, which is not the case for components from CLV.

The component-wise gradient boosting, or L2-boosting approach \citep{Buhl07,Hof14} has also been considered for comparison purposes using the \texttt{glmboost} function included in the R package \texttt{mboost}. We used the same cross-validation structure as previously and tested step length values $\nu=$ 0.01 and 0.1 to 1 by step of 0.1, with a maximal number of iteration of 1000. The best combination of the shrinkage parameter $\nu$ and the number of iterations, was chosen according to the $RMSE_{CV}$ criterion. For this example, the best combination was $\nu=0.3$ and a stopping iteration number $m_{stop}=30$, leading to a $RMSE_{CV}$ value of 2.45, which is similar to the optimal values observed with previous methods. With both of these parameters, 23 variables were selected: seven from $\mathcal{G}_1$, the five variables from $\mathcal{G}_2$, six variables from $\mathcal{G}_3$, and two variables from each of the last two group.

\subsection{Orange juice authentication case study} \label{Sec.authen}

In this case study, a $^1$H NMR spectroscopic profiling approach was investigated to discriminate between authentic and adulterated juices  \citep{Vig12}. In this study, we considered the adulteration of orange juice (\textit{Citrus sinensis}) with clementine juice (\textit{Citrus reticulata}). Supplementation of substitution with cheaper or easier to find similar fruit is one of the type of fraud conducted within the fruit juice industry. For the experiment, twenty pure orange juices and ten pure clementine juices were selected from the Eurofins database. They are deemed to be representative of the variability of the fruit juices available on the market. From these juices, 120 blends were prepared by mixing one of the twenty orange juices and one of the ten clementine juices in known proportions. The proportion of the clementine juice in a mix were 10\%, 20\%, $\ldots$, or 60\%. The experimental design is described in more detail in \citet{Vig12}. The database was completed with the twenty authentic orange juices, for a total of 140 juice samples, and these were analyzed on an $^1$H Nuclear Magnetic Resonance (NMR) spectroscopy platform.

The NMR variables are associated with chemical shifts, given in ppm. In the following, two spectral regions were simultaneously considered: The region from 6 to 9 ppm and the region from 0.5 to 2.3 ppm. The first spectral region mostly includes chemical shifts associated with aromatic components and the second one due to amino-acid-specific spectral shifts (among others). Between these regions lies the typical $^1$H NMR spectra for orange juice sugars  \citep{Rinke07}, which cannot discriminate between \textit{Citrus sinensis} and \textit{Citrus reticulata} juice. After a preprocessing binning step, 300 chemical shift variables were collected between 6 to 9 ppm,  and 180 variables between 0.5 to 2.3 ppm. 

\begin{figure}[h]
	\centering
	\resizebox*{6.25cm}{5cm}{\includegraphics{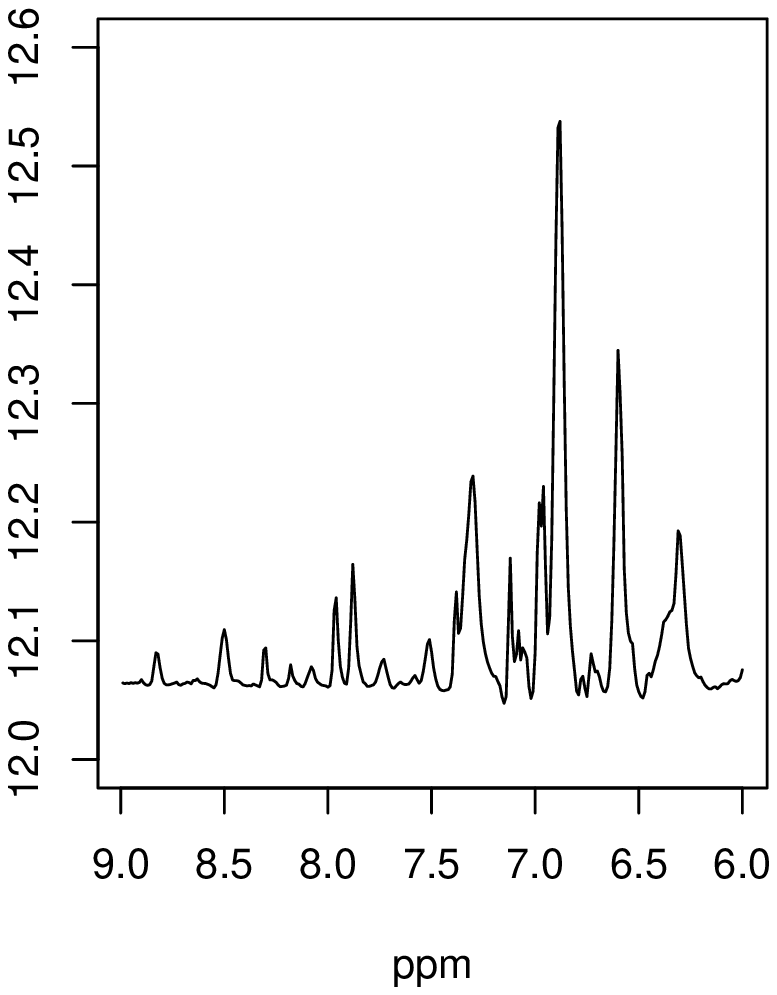}}
	\hspace{5pt}
	\resizebox*{3.75cm}{5cm}{\includegraphics{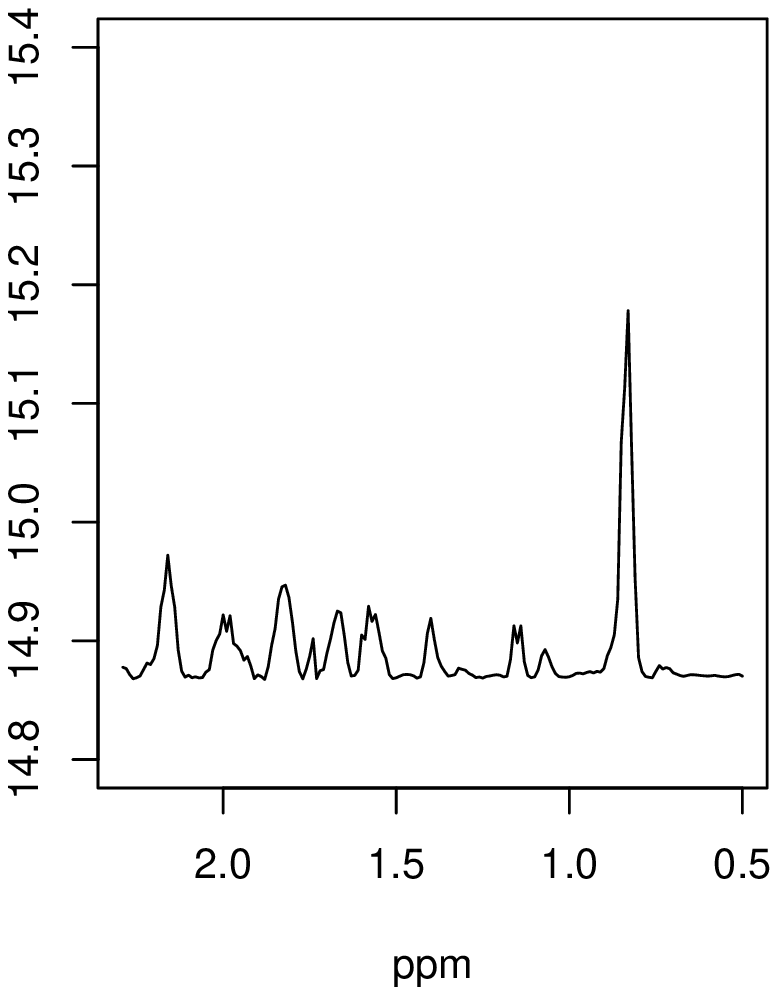}}
	\caption{Mean spectrum of the log-transformed NMR signals according to the spectral region. Two subplots are shown due to the differences with order of magnitude in both regions.} 
	\label{Fig:authen_mean_sprectrum}
\end{figure}

The data matrix is therefore composed of $n$=140 observations and $p$=480 variables. The log-transformed data are available in the R package \texttt{ClustVarLV} (dataset \texttt{authen\_NMR}). The mean of the log-transformed signals for both spectral regions are shown in Figure \ref{Fig:authen_mean_sprectrum}.

In the following, a pareto-scaling was applied to each variable. This scaling, which consists of dividing each variable by the square root of its standard deviation, was shown in this case study \citep{Vig12} to be preferable to the usual pre-scaling by the standard deviation. Moreover, the splitting of the observations into ten segments was defined for Cross-Validation purposes. A proportional stratified allocation rule has been adopted in such a way that each segment contained two observations of each of the seven experimental levels, ranging from 0 to 60\% of co-fruit added to pure orange juices.  

\begin{figure}[h]
	\centering
	\includegraphics[height=10cm]{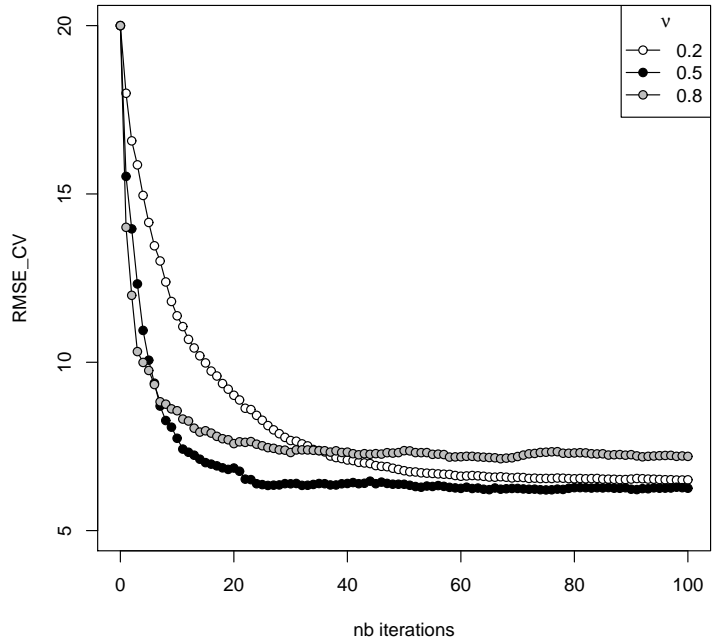}
	\caption{Evolution of the Root Mean Squared Error in Cross-validation with \textit{lmCLV}, according to the number of iterations, and the values of the shrinkage parameter, $\nu$, in the orange juice authentication case study.} 
	\label{Fig:authen_RMSECV}
\end{figure}

Figure \ref{Fig:authen_RMSECV} shows the evolution of the errors in prediction, evaluated by means of the $RMSE_{CV}$ criterion, when the shrinkage parameter $\nu$ was set to 0.2, 0.5 or 0.8. On this basis, we choose to consider a value of the shrinkage parameter $\nu=0.5$ for more detailed analysis of the model.

Besides the predictive purpose, one of the key features of \textit{lmCLV} is to identify groups of explanatory variables and their associated latent variables that are, step by step, selected and involved in the model. The introduction sequence of the CLV latent variables provides an interesting insight. However, it should be noted that the smaller the shrinkage parameter, the more often the same group of explanatory variables, by the means of its latent variable, will appear. Therefore, some of the \textit{lmCLV} algorithm's outputs are provided for each exhibited group, in order of its first occurrence, rather than by iteration. Thus, the OLS coefficients, $\mathbf{\alpha}^{(m)}$, associated with a selected latent component $\mathbf{c}^{*(m)}$ (see step 4 in Sect.\ref{Sec:model}) are aggregated on all iterations, $m$, to which this specific latent component is selected, up to the predefined maximum number of iterations, $M$. The same applies to the $\mathbf{\beta}$ coefficients of each of the (pre-processed) explanatory variables belonging to a selected group. Finally, a group importance criterion has been introduced. This importance is assessed as the sum, over all its occurrences, of the decrease in residual variance allowed by introducing the shrunken OLS estimate of the associated latent component.

\begin{figure}[h]
	\centering
	\includegraphics[width=\linewidth]{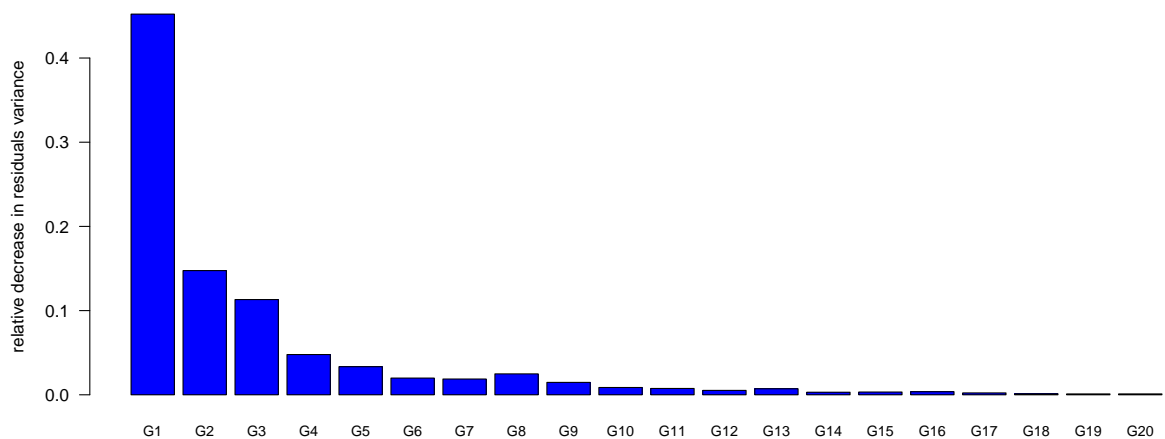}
	\caption{Relative Group Importance for the orange juice authentication case study. The Group Importance values are expressed relatively to the total variance of the response variable $\mathbf{y}$.} 
	\label{Fig:authen_GroupImp}
\end{figure}

In our case study, using the whole data set, with \textit{lmCLV} parameters $\nu=0.5$ and $M=25$, the group importance estimates are depicted in Figure \ref{Fig:authen_GroupImp}. This reveals the presence of a group of spectral variables (G1) of high importance, as well as two other important groups (G2 and G3). Each group is numbered according to its first occurrence, which corresponds rather well with its importance in the model. 

\begin{figure}[h]
	\centering
	\includegraphics[height=10cm]{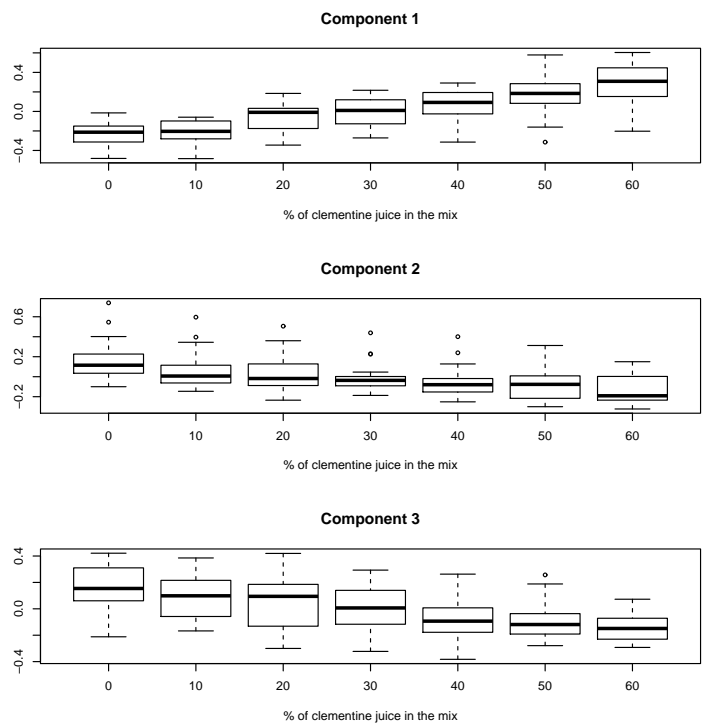}
	\caption{relationship between the first three group components and the proportion of clementine juice added to a mix.} 
	\label{Fig:authen_comp}
\end{figure}

The first group involved nine spectral variables at 7.52, 7.51, 7.50, 6.77, 2.10, 2.08, 2.07, 2.06 and 2.04 ppm. The spectral range between 7.50 and 7.52 ppm, combined with the signal at 6.77 ppm and around 2 ppm, is particularly interesting and could be associated with 4 amino-3-methylbenzoic acid (see for instance the Spectral Database for Organic Compounds \citep{SDBS}). This information is quite stable: the spectral variables at 7.52, 7.51, 6.77 and 2.08 ppm had been selected in the first group at each of the 10 iterations of the CV procedure. The same variables have also be noted in \citet{Vig12}, but their presumed association with the same compound could not be so clearly highlighted. The second group of spectral variables consisted of ten variables (7.15, 7.10-7.09, 1.10-1.05 and 0.96 ppm). The CV procedure showed that the range between 1.09 and 1.05 ppm and the peak at 7.10-7.09 ppm was simultaneously retrieved 7 times out of 10 as the second or third group. Lastly, the third group consisted of nine spectral variables, and specifically a range between 1.65 and 1.60 ppm. The relationship between these first three group components and the proportion of clementine juice in a mix is shown in Figure \ref{Fig:authen_comp}.

\medskip

In Table \ref{Tab:comparRMSECV}, the predictive ability of \textit{lmCLV} with $\nu=0.5$ is compared with those observed with Sparse PLS Regression \citep{Chun10}, Elastic-Net \citep{Zou05}, L2-boosting \citep{Buhl07,Hof14} and Random Forest \citep{Breiman01}. The $RMSE_{CV}$ were evaluated using the same 10-fold Cross-Validation division as above. For Sparse PLS Regression (sPLSR) the R package \texttt{spls} was used, with parameter $\eta \in \{0, 0.1, \ldots, 0.9\}$ and $K$, the number of components, varying from 1 to 20. Elastic-Net (E-Net) was adjusted using the R package \texttt{glmnet} and by considering the Gaussian family model with mixing parameter $\alpha$ fixed to 0.5. The parameter $\lambda$ was chosen as the mean of optimal values determined for each CV fold. For the L2-boosting approach (L2-boost), in the R package \texttt{mboost}, values of parameter $\nu$ from 0 (0.01 precisely) to 1, by 0.1, with a maximal number of 1000 iterations, was explored. RandomForest was considered because this machine learning approach often proves to be effective in high-dimensional classification or regression problems and provides an interesting variables importance ranking.  The R package \texttt{randomForest} was used with parameter $ntree=5000$, and default values for the two other parameters, $mtry$ (i.e. 160) and $nodesize$ (i.e. 5).

\begin{table}[h]
	\caption{\label{list}Root Mean Squared Error in cross-validated prediction of the percentage of clementine juice added to the orange juices, according to various methods.}
	{\begin{tabular}{llcc} 
			\hline
			method\textsuperscript{a} & parameters setting & $RMSE_{CV}$ & \# var. selected \\
			\hline
			lmCLV & $\nu=0.5$, $M=25$ & 6.366 & 145\\
			sPLSR & $\eta=0.3$, $K=5$ & 6.427 & 150\\ 
			E-Net & $\alpha=0.5$, $\lambda=0.08$ & 6.335 & 90 \\
			L2-boost & $\nu=0.2$, $M=209$ & 6.742 & 75\\
			RdF & $ntree=5000, mtry=160, nodesize=5$ &  9.992 & 146\textsuperscript{b} \\
			\hline
	\end{tabular}}
	\label{Tab:comparRMSECV}
	
	{\scriptsize 	\textsuperscript{a} All the methods was applied using available R packages. Their paramaters were determined on the basis of Cross-Validation with the same samples division. \\	\textsuperscript{b} For Random Forest, variables with a standardized variable importance value greater than 3 were selected.}
\end{table}

As shown in Table \ref{Tab:comparRMSECV}, \textit{lmCLV} has an expected prediction performance similar to Sparse PLS Regression, Elastic-Net and the L2-boosting approach. Quite surprisingly, in this case study, Random Forest did not perform very well. In fact, one can observe that by Random Forest the prediction of the percentage of co-fruit added was overestimated or underestimated at the extremes of the experimental scale, i.e. for 0-10\% or 50-60\%.

\begin{figure}[h]
	\centering
	\includegraphics[width=\linewidth]{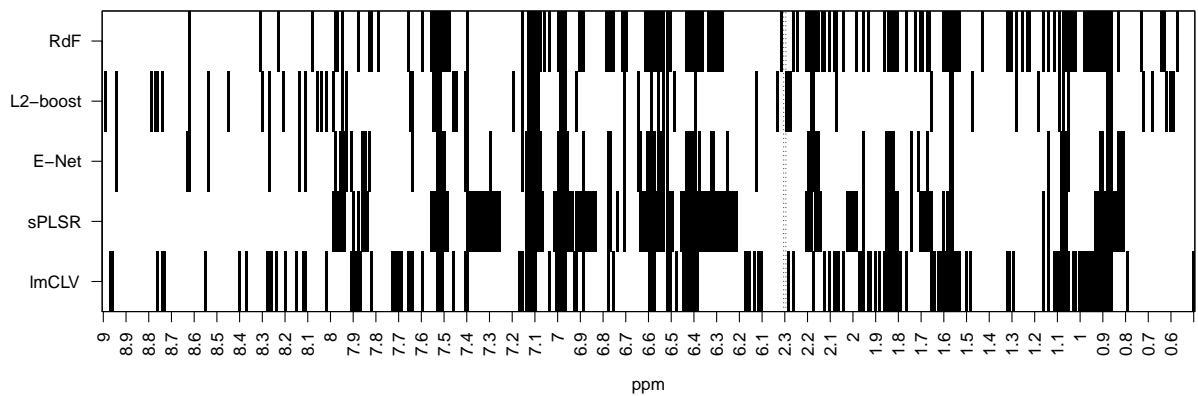}
	\caption{Location of the spectral variables involved in the models according to the method considered, in the orange juice authentication case study. } 
	\label{Fig:authen_varsel}
\end{figure}

The number of variables involved in the model fitted on the whole dataset, using the predetermined model's parameters, is indicated in the last column of Table \ref{Tab:comparRMSECV}. As shown in Figure \ref{Fig:authen_varsel}, while Sparse PLS Regression has the tendency to identify fewer but larger spectral ranges, the L2-boosting approach retained a relatively small number of variables associated with narrow ranges. For \textit{lmCLV}, 20 groups and their latent variable were identified. Since several small spectral ranges were merged within the same group, the total number of spectral variables involved was rather high. However, the order of extraction and the grouping effect, which are a specificity of \textit{lmCLV}, cannot be revealed in Figure \ref{Fig:authen_varsel}. Globally, 19 spectral variables were retained with the five methods considered herein. We systematically found the variables at 7.51-7.52 ppm, but not the other variables that belonged to the first group extracted with \textit{lmCLV}. The variables between 7.09 to 7.12 ppm as well as the variable at 1.07 ppm are also present, as in the second group extracted with \textit{lmCLV}. We finally noted that the areas at 6.96-6.98 ppm, 1.56-1.57 ppm, 0.86-0.88 ppm were selected with the five methods.

\section{Conclusion}
In this study, we introduced a strategy for linear model fitting based on the hypothesis that high-dimensional datasets often include highly correlated variables having similar effect on the response variable. This is specifically the case for modern scanning instruments such as those used in spectroscopy (infrared, near-infrared, Raman, NMR,...) which are able to collect a large number of sequential spectral variables, several of them being representative of one feature/signal. Omic data, and specifically metabolomic data, contain a large quantity of measured elements that are components of the same metabolic pathways and that constitute the biological information that is sought. The basis of the approach is to then cluster the explanatory variables first, as other authors have already proposed \citep{Buhl13,Enki13}. In \textit{lmCLV}, the clustering stage is based on the CLV method and consists of identifying unidimensional latent components which represent clusters of variables. These latent components play the role of the predictors in the second stage which results in introducing the explanatory variables in a group-wise fashion.

In constrast with \citet{Buhl13} for the CRL approach (Cluster Representative Lasso), and \citet{Hastie01} or \citet{Park07} in which each cluster representative is defined as the mean variable of a group of variables, the CLV latent components are defined at the same time as the clusters, and are derived from the eigen decomposition of the within cluster covariance matrix. In addition, for the construction of the regression model itself, we have adopted an L2-boosting approach, rather than a Lasso approach. This makes it possible to build a simple and efficient algorithm in which the CLV dendrogram constitutes the base-learner. 

Compared to our previous study \citep{Chen16}, the algorithm proposed here requires much less computer resources. In the previous study, the clustering stage was performed on the data matrix combining the residuals of the response variable and the explanatory variables, and was repeated for each iteration. However, the  clustering of several hundred variables requires the largest part of the computation time. Using the new version of \textit{lmCLV}, for the case study on orange juice authentication (Sect. \ref{Sec.authen}), which involved 480 explanatory variables, the whole procedure ($\nu=0.5$, $M=100$) took 1 min 48 sec on one 3.4 GHz processor, including 1 min 25 sec for the clustering stage alone. For the implementation of this algorithm, a function \texttt{lmCLV} will be available in version 2.0.2 of the R package \texttt{ClustVarLV}.

In both examples presented in this study, \textit{lmCLV} was shown to have similar predictive efficiency to other methods, and importantly, provided additional interpretability capacity. The use of latent components that are easy to interpret, because each of them is associated with a group of collinear variables, is consistent with the development of  modern simplifying approaches for modelization. However, compared to Lasso-based methodologies, dimensionality reduction based on variables clustering  makes it easier to identify directions of interest for prediction and makes it possible to highlight functional links between explanatory variables where they exist.

\section*{Acknowledgements}

The author is grateful to Freddy Thomas (Authenticity unit, Eurofins Analytics France, Nantes, France) for the opportunity to present the case study on orange juice authentication.

\bibliographystyle{chicago}
\bibliography{lmCLV}            

\end{document}